\documentclass{article}%
\usepackage{amsfonts}
\usepackage{amsmath}
\usepackage{amssymb}
\usepackage{graphicx}%
\providecommand{\U}[1]{\protect\rule{.1in}{.1in}}

\begin{document}

\title{Trading drift and fluctuations in entropic dynamics: quantum dynamics as an
emergent universality class\thanks{Invited paper presented at the EmQM15
Workshop on Emergent Quantum Mechanics, Vienna University of Technology
(October 23--25, 2015). }}
\author{Daniel Bartolomeo and Ariel Caticha\\{\small Department of Physics, University at Albany--SUNY, Albany, NY 12222,
USA}}
\date{}
\maketitle

\begin{abstract}
Entropic Dynamics (ED) is a framework that allows the formulation of dynamical
theories as an application of entropic methods of inference. In the generic
application of ED to derive the Schr\"{o}dinger equation for $N$ particles the
dynamics is a non-dissipative diffusion in which the system follows a
\textquotedblleft Brownian\textquotedblright\ trajectory with fluctuations
superposed on a smooth drift. We show that there is a family of ED models that
differ at the \textquotedblleft microscopic\textquotedblright\ or sub-quantum
level in that one can enhance or suppress the fluctuations relative to the
drift. Nevertheless, members of this family belong to the same universality
class in that they all lead to the same emergent Schr\"{o}dinger behavior at
the \textquotedblleft macroscopic\textquotedblright\ or quantum level. The
model in which fluctuations are totally suppressed is of particular interest:
the system evolves along the smooth lines of probability flow. Thus ED
includes the Bohmian or causal form of quantum mechanics as a special limiting
case. We briefly explore a different universality class -- a non-dissipative
dynamics with microscopic fluctuations but no quantum potential. The Bohmian
limit of these hybrid models is equivalent to classical mechanics. Finally we
show that the Heisenberg uncertainty relation is unaffected either by
enhancing or suppressing microscopic fluctuations or by switching off the
quantum potential.

\end{abstract}

\section{Introduction}

Entropic Dynamics (ED) is a framework in which quantum theory is derived as an
application of entropic methods of inference.\footnote{The principle of
maximum entropy as a method for inference can be traced to the pioneering work
of E. T. Jaynes. For a pedagogical overview of Bayesian and entropic inference
and further references see \cite{Caticha 2012}.} As in any theory of
inference, establishing the subject matter is the first and most crucial step;
this amounts to a choice of microstates, that is, a choice of \emph{beables}.
Once that choice is made the dynamics is driven by entropy subject to
constraints which reflect the information needed for making physical
predictions \cite{Caticha 2010a}-\cite{Caticha 2015}.

ED naturally leads to an epistemic view of the quantum state $\psi$ but with
an added twist. Within an inferential framework it is not sufficient to merely
interpret the probability $|\psi|^{2}$ as a state of knowledge.\footnote{For a
review with references on the epistemic vs ontic interpretations of the
quantum state see \cite{Leifer 2014}.} It is just as important to require that
the dynamics, that is the updates of $\psi$, be consistent with the
established rules for updating probability distributions. This is where the
method of maximum entropy and the Bayesian methods enter. Furthermore, the
entropic dynamics must include \emph{both} the unitary time evolution
described by the Schr\"{o}dinger equation \emph{and} the collapse of the wave
function during measurement. As a result the ED framework turns out to be very
restrictive. In a fully \emph{entropic} dynamics we do not postulate an
underlying mechanics with an action principle that operates at some deeper
level. Instead, at the sub-quantum level there is only inference and at the
quantum level the emergent dynamics is described by an action principle that
is derived rather than posited.

There is a vast literature on the attempts to reconstruct quantum mechanics
and it is inevitable that the ED approach will resemble them in one aspect or
another. Indeed, to the extent that any of these approaches are successful
they must sooner or later converge to the same Schr\"{o}dinger equation.
However, there are important differences. For example, the central concern
with the notion of time makes ED significantly different from other approaches
that are also based on information theory (such as \emph{e.g.}, \cite{Wootters
1981}-\cite{Reginatto 2013}). And ED also differs from those approaches (see
\emph{e.g.}, \cite{Nelson 1985}-\cite{Grossing 2008}) that aim to explain the
emergence of quantum behavior as the effective statistical mechanics of some
underlying sub-quantum mechanics which might possibly include some additional
stochastic element. Indeed, ED makes no reference to any sub-quantum action
principles whether classical, deterministic, or stochastic.

As stated above in ED inferences are carried out on the basis of information
introduced in the form of constraints. In the particular case of the ED of $N$
particles the microstates are the positions of the particles. The basic
physical input --- that the particles follow continuous trajectories --- is
implemented through $N$ constraints, one for each particle. The multiple roles
played by the corresponding Lagrange multipliers $\alpha_{n}$ ($n=1\ldots N$)
are by now well understood. The multipliers regulate the flow of time and they
serve to unify the concept of mass with that of quantum fluctuations.

If these were the only constraints each particle would experience its own
independent dynamics and the resulting motion would be an isotropic diffusion.
It turns our that one can obtain more interesting forms of dynamics by
imposing just one single additional constraint. This constraint acts on
configuration space and involves a \textquotedblleft
drift\ potential,\textquotedblright\ an epistemic tool that plays a role
somewhat analogous to that of a pilot wave. The drift potential contributes to
the phase of the wave function, it correlates the motion of particles, and it
causes such quintessential quantum effects as interference and entanglement.
The corresponding multiplier $\alpha^{\prime}$ is not nearly as well
understood and the purpose of this paper is to fill this gap \cite{Bartolomeo
Caticha 2015}.

We begin with a brief overview of ED following the presentation in
\cite{Caticha 2014b}. We show that the role of the multiplier $\alpha^{\prime
}$ is to control the relative magnitudes of drift and fluctuations. Our main
result is simple: ED models with different values of $\alpha^{\prime}$ lead to
the same Schr\"{o}dinger equation. In other words, different \textquotedblleft
microscopic\textquotedblright\ or sub-quantum models can lead to the same
emergent quantum behavior --- they belong to the same \textquotedblleft
universality\textquotedblright\ class. The limit of large $\alpha^{\prime}$
deserves particular attention. In this limit fluctuations are suppressed, the
drift motion prevails, and the particles tend to move along the smooth lines
of probability flow. This means that ED includes the Bohmian form of quantum
mechanics as a special limiting case \cite{Bohm 1952}-\cite{Holland 1993}.

We briefly explore one alternative family of ED models --- a different
universality class. At the microscopic level these models also describe
particles that follow Brownian trajectories but without the non-local
correlations induced by the quantum potential \cite{Nawaz Caticha 2011}. The
multiplier $\alpha^{\prime}$ also acts to suppress microscopic fluctuations
and the Bohmian, or large $\alpha^{\prime}$ limit, approaches classical
mechanics. All these microscopic models lead to the same dynamics at the
macroscopic level --- the emergent dynamics is an essentially classical mechanics.

The value of the ED approach to quantum theory lies in part in the conceptual
clarity it brings to issues of interpretation. In \cite{Nawaz Caticha 2011}
the methods originally developed in the context of stochastic mechanics
\cite{de Falco et al 1982}-\cite{Golin 1986} were adapted to derive the
Heisenberg uncertainty relation for position and momentum within the context
of ED. The fact that in ED we can enhance or suppress microscopic fluctuations
relative to the drift and that we can switch on or off the quantum potential
leads us to raise the question of how these changes affect the uncertainty
relations. Are there potential violations of the uncertainty principle? We
find that the uncertainty relations are unaffected by changes in the
microscopic fluctuations, even when suppressing them to the Bohmian limit, or
by switching off the quantum potential.

\section{Entropic Dynamics --- a brief overview}

We consider the ED of $N$ particles living in a flat Euclidean space
$\mathbf{X}$ with metric $\delta_{ab}$. In the ED framework the particles have
definite positions $x_{n}^{a}$ and it is their unknown values that we wish to
infer. (The index $n$ $=1\ldots N$ labels the particle and $a=1,2,3$ its
spatial coordinates.) The position of the system in configuration space
$\mathbf{X}_{N}=\mathbf{X}\times\ldots\times\mathbf{X}$ will also be denoted
$x^{A}$ where $A=(n,a)$.

The main dynamical assumption is that motion is continuous which means that it
can be analyzed as a sequence of many infinitesimally short steps. Thus we
first find the probability $P(x^{\prime}|x)$ that the system takes a short
step from $x^{A}$ to $x^{\prime A}=x^{A}+\Delta x^{A}$ and then we determine
how such short steps accumulate. To find $P(x^{\prime}|x)$ we maximize the
(relative) entropy,
\begin{equation}
\mathcal{S}[P,Q]=-\int dx^{\prime}\,P(x^{\prime}|x)\log\frac{P(x^{\prime}%
|x)}{Q(x^{\prime}|x)}~,\label{Sppi}%
\end{equation}
where we adopt the notation $dx^{\prime}=d^{3N}x^{\prime}$. The prior
distribution $Q(x^{\prime}|x)$ describes the state of knowledge (of an ideally
rational agent) \emph{before} any information about the motion is taken into
account. We shall assume that $Q(x^{\prime}|x)$ reflects extreme ignorance
which is expressed by a uniform distribution: $Q(x^{\prime}|x)dx^{\prime}$ is
proportional to the volume element.\footnote{Strictly uniform non-normalizable
priors are mathematically problematic. This difficulty can be avoided by
adopting a physically reasonable normalizable prior. By \textquotedblleft
uniform\textquotedblright\ we actually mean any distribution that is
essentially flat over macroscopic scales.} Since the space $\mathbf{X}_{N}$ is
flat we can set $Q(x^{\prime}|x)=1$.

The physical information about the motion is introduced through constraints.
The fact that particles take infinitesimally short steps from $x_{n}^{a}$ to
$x_{n}^{\prime a}=x_{n}^{a}+\Delta x_{n}^{a}$ is imposed through $N$ separate
constraints,
\begin{equation}
\langle\Delta x_{n}^{a}\Delta x_{n}^{b}\rangle\delta_{ab}=\kappa_{n}%
~,\qquad(n=1\ldots N)~ \label{kappa n}%
\end{equation}
where $\kappa_{n}$ are constants. The $\kappa_{n}$'s are chosen to be constant
to reflect the translational symmetry of the space $\mathbf{X}$; they are
$n$-dependent in order to accommodate non-identical particles; and eventually
we take $\kappa_{n}\rightarrow0$ to implement infinitesimally short steps.

There is one additional constraint that leads to correlations among the
particles,
\begin{equation}
\langle\Delta x^{A}\rangle\partial_{A}\phi=\sum\limits_{n=1}^{N}\left\langle
\Delta x_{n}^{a}\right\rangle \frac{\partial\phi}{\partial x_{n}^{a}}%
=\kappa^{\prime}~, \label{kappa prime}%
\end{equation}
which introduces the \textquotedblleft drift\textquotedblright\ potential
$\phi$ and $\partial_{A}=\partial/\partial x^{A}=\partial/\partial x_{n}^{a}$.
$\kappa^{\prime}$ is another small but for now unspecified
position-independent constant. Eq.(\ref{kappa prime}) is a single constraint
that acts on the $3N$-dimensional configuration space.\footnote{Elsewhere, in
the context of particles with spin, we will see that the potential $\phi(x)$
can be given a natural geometric interpretation as an angular variable. Its
integral over any closed loop is $%
{\displaystyle\oint}
d\phi=2\pi n$ where $n$ is an integer.}

Maximizing the entropy (\ref{Sppi}) subject to (\ref{kappa n}),
(\ref{kappa prime}), and normalization leads to
\begin{equation}
P(x^{\prime}|x)=\frac{1}{\zeta}\exp[-\sum_{n}(\frac{1}{2}\alpha_{n}\,\Delta
x_{n}^{a}\Delta x_{n}^{b}\delta_{ab}-\alpha^{\prime}\Delta x_{n}^{a}%
\frac{\partial\phi}{\partial x_{n}^{a}})]~, \label{Prob xp/x a}%
\end{equation}
where $\zeta$ is a normalization constant and $\alpha_{n}$ and $\alpha
^{\prime}$ are the Lagrange multipliers associated to (\ref{kappa n}) and
(\ref{kappa prime}). The limit of infinitesimally short steps $\kappa
_{n}\rightarrow0$ is achieved as $\alpha_{n}\rightarrow\infty$.

Already at this early stage we can see that ED\ exhibits a rather trivial
symmetry: Imposing the constraint (\ref{kappa prime}) with the pair
$(\phi,\kappa^{\prime})$ leads to the same transition probability
$P(x^{\prime}|x)$ as a constraint with the pair $(C\phi,C\kappa^{\prime})$
where $C$ is some arbitrary constant. In previous work \cite{Caticha 2014b} we
took advantage of the symmetry and rescaled with $C=1/\alpha^{\prime}$. This
amounts to setting $\alpha^{\prime}\phi\rightarrow\phi$ which eliminates
$\alpha^{\prime}$. Here we wish to examine the effect of $\alpha^{\prime}$ for
given $\phi$ so we will keep it explicit.

To find how these short steps accumulate to produce a finite change we
introduce a book-keeping device --- this is how the notion of time enters
dynamics. As discussed in \cite{Caticha 2010a}-\cite{Caticha 2015} entropic
time is measured by the fluctuations themselves (see eq.(\ref{fluc}) below)
which leads to the choice
\begin{equation}
\alpha_{n}=\frac{m_{n}}{\eta\Delta t}~, \label{alpha n}%
\end{equation}
where $\Delta t$ is the interval over which the short step $x\rightarrow
x^{\prime}$ occurs, the $m_{n}$ are particle-specific constants that will be
called \textquotedblleft masses\textquotedblright,\ and $\eta$ is a constant
that fixes the units of time relative to those of length and mass. With this
choice of $\alpha_{n}$ a generic displacement can be expressed as an expected
drift plus a fluctuation,
\begin{equation}
\Delta x^{A}=b^{A}\Delta t+\Delta w^{A}~, \label{Delta x}%
\end{equation}
where $b^{A}(x)$ is the drift velocity,
\begin{equation}
\langle\Delta x^{A}\rangle=b^{A}\Delta t\quad\text{with}\quad b^{A}=\frac
{\eta\alpha^{\prime}}{m_{n}}\delta^{AB}\partial_{B}\phi=\eta\alpha^{\prime
}m^{AB}\partial_{B}\phi~. \label{drift velocity}%
\end{equation}
$m_{AB}=m_{n}\delta_{AB}$ is the \textquotedblleft mass\textquotedblright%
\ tensor and $m^{AB}=\delta^{AB}/m_{n}$ is its inverse. The fluctuations
$\Delta w^{A}$ satisfy,
\begin{equation}
\langle\Delta w^{A}\rangle=0\quad\text{and}\quad\langle\Delta w^{A}\Delta
w^{B}\rangle=\frac{\eta}{m_{n}}\delta^{AB}\Delta t=\eta m^{AB}\Delta t~.
\label{fluc}%
\end{equation}
Comparing equations (\ref{drift velocity}) and (\ref{fluc}) for short steps,
as $\Delta t\rightarrow0$, we see that the fluctuations are much larger than
the drift ($\Delta w^{A}\sim\Delta t^{1/2}$ while $\langle\Delta x^{A}%
\rangle\sim\Delta t$) which leads to non-differentiable trajectories
characteristic of a Brownian motion. They also show that for fixed $\phi$ the
effect of the multiplier $\alpha^{\prime}$ is to enhance or suppress the drift
$b^{A}\Delta t$ relative to the fluctuations $\Delta w^{A}$.

Having introduced a convenient notion of time through (\ref{alpha n}), the
accumulation of many short steps leads to a probability distribution
$\rho(x,t)$ in configuration space that obeys a Fokker-Planck equation (FP),
\cite{Caticha 2012}\cite{Caticha 2010a}\cite{Caticha 2014a}
\begin{equation}
\partial_{t}\rho=-\partial_{A}\left(  \rho v^{A}\right)  ~.\label{FP b}%
\end{equation}
In this equation $v^{A}$ is the velocity of the probability flow in
configuration space or \emph{current velocity}. It is given by
\begin{equation}
v^{A}=b^{A}+u^{A}\quad\text{where}\quad u^{A}=-\eta m^{AB}\partial_{B}\log
\rho^{1/2}~
\end{equation}
is called the \emph{osmotic velocity}. The interpretation of $u^{A}$ follows
immediately from looking at its contribution to the probability flux,
\begin{equation}
\rho u^{A}=-\frac{1}{2}\eta m^{AB}\partial_{B}\rho\text{~}.
\end{equation}
This is recognized as the analogue of Fick's law of diffusion with a diffusion
tensor $\eta m^{AB}/2$. Since both $b^{A}$ and $u^{A}$ are gradients the
current velocity $v^{A}$ is a gradient too,%
\begin{equation}
v^{A}=m^{AB}\partial_{B}\Phi\quad\text{where}\quad\Phi=\eta\alpha^{\prime}%
\phi-\eta\log\rho^{1/2}\label{curr}%
\end{equation}
will be called the \emph{phase}. Thus, the phase has a drift component and a
diffusive or osmotic component.

The dynamics described by the FP equation (\ref{FP b}) is a standard
diffusion. To describe a \textquotedblleft mechanics\textquotedblright\ we
require that the diffusion be \textquotedblleft
non-dissipative\textquotedblright\ which is achieved by an appropriate
readjustment of the constraint (\ref{kappa prime}) after each step $\Delta t$.
The net effect is that the drift potential $\phi$, or equivalently the phase
$\Phi$, is promoted to a fully dynamical degree of freedom. The diffusion is
said to be \textquotedblleft non-dissipative\textquotedblright\ when the
actual updating of $\Phi$ is implemented by imposing that a certain functional
$\tilde{H}[\rho,\Phi]$ be conserved. In order to offset the entropic change
$\rho\rightarrow\rho+\delta\rho$, one requires that $\Phi$ changes
$\Phi\rightarrow\Phi+\delta\Phi$ in such a way that
\begin{equation}
\tilde{H}[\rho+\delta\rho,\Phi+\delta\Phi]=\tilde{H}[\rho,\Phi]~.
\end{equation}
As shown in \cite{Caticha 2014b} the requirement that $\tilde{H}$ be conserved
for arbitrary choices of $\rho$ and $\Phi$ implies that the coupled evolution
of $\rho$ and $\Phi$ is given by a conjugate pair of Hamilton's equations,
\begin{equation}
\partial_{t}\rho=\frac{\delta\tilde{H}}{\delta\Phi}\quad\text{and}%
\quad\partial_{t}\Phi=-\frac{\delta\tilde{H}}{\delta\rho}~.\label{Hamilton}%
\end{equation}
The \textquotedblleft ensemble\textquotedblright\ Hamiltonian $\tilde{H}$ is
chosen so that the first equation in (\ref{Hamilton}) reproduces the FP
equation (\ref{FP b}). Then the second equation becomes a Hamilton-Jacobi
equation (HJ). Further arguments from information geometry can then be invoked
to fully specify the form of the functional $\tilde{H}[\rho,\Phi]$
\cite{Caticha 2014b}. They suggest that the natural choice of $\tilde{H}$ is%
\begin{equation}
\tilde{H}[\rho,\Phi]=\int dx\,\rho\left[  \frac{1}{2}m^{AB}\partial_{A}%
\Phi\partial_{B}\Phi+V+\xi m^{AB}\frac{1}{\rho^{2}}\partial_{A}\rho
\partial_{B}\rho\right]  ~.\label{Hamiltonian}%
\end{equation}
The first term in the integrand is the \textquotedblleft
kinetic\textquotedblright\ term that reproduces the FP equation (\ref{FP b}).
The second term represents the simplest non-trivial interaction, a potential
energy that is linear in $\rho$ and introduces the standard potential $V(x)$.
The third term, motivated by information geometry, is the trace of the Fisher
information and is called the \textquotedblleft quantum\textquotedblright%
\ potential. The parameter $\xi$\ turns out to be crucial: it controls the
relative contributions of the two potentials. When $\xi>0$ we write $\xi
=\hbar^{2}/8$. Thus $\xi$ defines the value of what we call Planck's constant
$\hbar$, and sets the scale that separates quantum from classical regimes. The
case $\xi=0$ will be addressed below; the case $\xi<0$ leads to instabilities
and will not be discussed further.

To conclude this brief review of ED we note that at this point the dynamics is
fully specified by equations (\ref{Hamilton}) and (\ref{Hamiltonian}). Nothing
prevents us however from combining $\rho$ and $\Phi$ into a single complex
function,
\begin{equation}
\Psi=\rho^{1/2}\exp(i\Phi/\hbar)~. \label{psi k}%
\end{equation}
Then the pair of Hamilton's equations (\ref{Hamilton}) can be tremendously
simplified and written as a single complex linear equation,%
\begin{equation}
i\hbar\partial_{t}\Psi=-\frac{\hbar^{2}}{2}m^{AB}\partial_{A}\partial_{B}%
\Psi+V\Psi~, \label{sch c}%
\end{equation}
which we recognize as the Schr\"{o}dinger equation.

\section{Trading drift and fluctuations}

According to ED at the microscopic sub-quantum level the dynamics is very
irregular. The particles perform a Brownian motion, eq.(\ref{Delta x}), with
an expected drift and fluctuations given by (\ref{drift velocity}) and
(\ref{fluc}). From these equations we see that the effect of $\alpha^{\prime}$
is to enhance the drift relative to the fluctuations. This means that
different values of the multiplier $\alpha^{\prime}$ correspond different
types of microscopic dynamics.

We can study the sub-quantum effect of $\alpha^{\prime}$ directly through
eqs.(\ref{drift velocity}) and (\ref{fluc}). However, it may be more
instructive to rescale $\eta$ and write $\eta=\tilde{\eta}/\alpha^{\prime}$.
Under such rescaling the $\alpha^{\prime}$ dependence migrates from the drift
to the fluctuations,
\begin{equation}
\langle\Delta x^{A}\rangle=\tilde{\eta}m^{AB}\partial_{B}\phi\,\Delta
t\quad\text{and}\quad\langle\Delta w^{A}\Delta w^{B}\rangle=\frac{\tilde{\eta
}}{\alpha^{\prime}}m^{AB}\Delta t~, \label{fluc b}%
\end{equation}
and we see that increasing $\alpha^{\prime}$ at fixed $\tilde{\eta}$ has the
effect of suppressing the fluctuations.

In contrast, at the \textquotedblleft macroscopic\textquotedblright\ or
quantum level the dynamics is very smooth. It is a non-dissipative diffusion
described by Hamilton's equations (\ref{Hamilton}) or by the Schr\"{o}dinger
equation (\ref{sch c}). The quantum dynamics is clearly independent of
$\alpha^{\prime}$ which means that there is a whole family of microscopic
models --- one could call it a universality class --- that lead to the same
emergent quantum dynamics.\ {}

From eq.(\ref{curr}) we have
\begin{equation}
v^{A}=m^{AB}\partial_{B}\Phi\quad\text{with}\quad\Phi=\tilde{\eta}\phi
-\frac{\tilde{\eta}}{\alpha^{\prime}}\log\rho^{1/2}~. \label{phase}%
\end{equation}
Here too we see that when we change $\alpha^{\prime}$ the phase $\Phi$ remains
unchanged but the relative contributions of the drift and osmotic components change.

\paragraph{The Bohmian limit}

For large $\alpha^{\prime}$ the fluctuations are suppressed; osmotic or
diffusion effects become negligible. As $\alpha^{\prime}$ increases the
particles follow smoother trajectories which resemble a Brownian motion only
at increasingly shorter spatial scales. In the limit $\alpha^{\prime
}\rightarrow\infty$ we have
\begin{equation}
\Phi\rightarrow\tilde{\eta}\phi\quad\text{so that}\quad v^{A}\rightarrow
b^{A}~.
\end{equation}
The current and the drift velocities coincide and particles follow smooth
trajectories that coincide with the lines of probability flow. This is exactly
the kind of motion postulated by Bohmian mechanics \cite{Bohm 1952}%
-\cite{Holland 1993}.

But a word of caution is necessary. ED is driven by entropy; it is essentially
non-causal and indeterministic. This is what led us to introduce probabilities
and allowed us to write down a Fokker-Planck equation. Therefore the $\Delta
t\rightarrow0$ limit or, going back to (\ref{alpha n}), the $\alpha
_{n}\rightarrow\infty~$limit, which is the limit that enforces the continuity
of trajectories, must be taken at fixed $\alpha^{\prime}$, before we consider
the effect of $\alpha^{\prime}\rightarrow\infty$. Only then, once we recognize
that we are dealing with a tricky singular limit, can we claim that entropic
dynamics includes Bohmian mechanics as a special limiting case. Thus, no
matter how large the (fixed) value of $\alpha^{\prime}$, entropic dynamics
remains \textquotedblleft entropic\textquotedblright. Even for large
$\alpha^{\prime}$ the dynamics is still driven by fluctuations and at
sufficiently microscopic scales the expected motion is Brownian.

We must also emphasize that it is only with respect to the mathematical
formalism that ED includes Bohmian mechanics as a special case. The
philosophical differences constitute an unbridgeable gap. Bohmian mechanics
attempts to provide an actual description of reality, a description of the
ontology of the universe as it \textquotedblleft really\textquotedblright\ is
and as it \textquotedblleft really\textquotedblright\ happens. In the Bohmian
view the universe consists of real particles that have definite positions and
their trajectories are guided by a real field, the wave function $\Psi$
\cite{Bohm 1952}.

On the other hand ED is a purely epistemic theory. Its pragmatic goal is less
ambitious: to make the best possible predictions on the basis of very
incomplete information. In ED the particles also have definite positions and
its formalism includes a function $\Phi$ that behaves as a wave. But $\Phi$ is
a tool for reasoning; it is not meant to represent anything real. There is no
implication that the particles move the way they do because they are pushed
around by a pilot wave or by some stochastic force. In fact ED is silent on
the issue of what causative power is responsible for the peculiar motion of
the particles. What the probability $\rho$ and the phase $\Phi$ are designed
to do is not to guide the particles but to guide our inferences. They guide
our expectations of where and when to find the particles but they do not exert
any causal influence on the particles themselves.

\section{Another universality class and its Bohmian limit}

ED as a non-dissipative diffusion is defined by Hamilton's equations
(\ref{Hamilton}). The ensemble Hamiltonian, eq.(\ref{Hamiltonian}), includes a
parameter $\xi$ that regulates the strength of the quantum potential. Any
non-zero value $\xi>0$ yields a fully \emph{quantum} mechanics, albeit with
differing values of $\hbar$. We can also treat $\xi$ and $\hbar$ as
independent parameters and set $\xi=0$.\footnote{In this model $\xi$ and
$\hbar$ are independent parameters. $\xi$ is set to $0$ and $\hbar$ is defined
as the constant with the appropriate units of action that is needed to define
a wave function $\Psi=\rho^{1/2}e^{i\Phi/\hbar}$.} This leads us to a
qualitatively different theory --- a different universality class.

At the microscopic level the particles follow irregular Brownian trajectories
described by eqs.(\ref{Delta x}-\ref{fluc}). For $\xi=0$, $\alpha^{\prime}$
has the same effect of enhancing the drift relative to the fluctuations, and
therefore different values of $\alpha^{\prime}$ correspond to different types
of microscopic dynamics.

At the \textquotedblleft macroscopic\textquotedblright\ level the emergent
behavior is smooth. According to equations (\ref{Hamilton}) and
(\ref{Hamiltonian}) for $\xi=0$ the probability $\rho$ follows the gradient of
$\Phi$,
\begin{equation}
\partial_{t}\rho=\frac{\delta\tilde{H}}{\delta\Phi}=-\partial_{A}\left(  \rho
v^{A}\right)  \quad\text{with}\quad v^{A}=m^{AB}\partial_{B}\Phi~,
\end{equation}
and $\Phi$ evolves according to
\begin{equation}
\partial_{t}\Phi=-\frac{\delta\tilde{H}}{\delta\rho}=-\frac{1}{2}%
m^{AB}\partial_{A}\Phi\partial_{B}\Phi-V~, \label{classical HJ}%
\end{equation}
which we recognize as the classical Hamilton-Jacobi equation. Therefore the
probability $\rho$ flows along the classical path. In fact, these are exactly
the classical equations of motion in a Liouville representation. We conclude
that the emergent macroscopic dynamics is essentially classical mechanics.
There is, however, no implication that the particles themselves follow the
classical paths. Indeed, at any instant of time the particles undergo the same
fluctuations, eq.(\ref{fluc b}), that we would expect for any non-zero value
of $\xi$.

For $\xi=0$ ED is a hybrid theory; it resembles classical mechanics in some
respects and quantum mechanics in others. Just as in quantum mechanics the
particles follow Brownian paths and the dynamics is a non-dissipative
diffusion; they even satisfy an uncertainty principle \cite{Nawaz Caticha
2011}. On the other hand, just as in classical mechanics, the probability
flows according to paths described by the classical Hamilton-Jacobi equation.
One can even combine $\rho$ and $\Phi$ into a single complex function,
$\Psi=\rho^{1/2}\exp(i\Phi/\hbar)$, and write the coupled evolution of $\rho$
and $\Phi$ in terms of a single complex equation that resembles a
Schr\"{o}dinger equation,
\begin{equation}
i\hbar\partial_{t}\Psi_{k}=-\frac{\hbar^{2}}{2}m^{AB}\partial_{A}\partial
_{B}\Psi_{k}+V\Psi_{k}+\frac{\hbar^{2}}{2}m^{AB}\frac{\partial_{A}\partial
_{B}|\Psi_{k}|}{|\Psi_{k}|}\Psi_{k}~.\label{sch b}%
\end{equation}
But this equation is not linear which means that a central feature of quantum
behavior, the superposition principle, has been lost.

Within the family of microscopic models with $\xi=0$ we can also take the
\textquotedblleft Bohmian\textquotedblright\ limit, $\alpha^{\prime
}\rightarrow\infty$. Increasing $\alpha^{\prime}$ at fixed $\tilde{\eta}$
suppresses the fluctuations so the particles follow smoother trajectories that
increasingly approximate the lines of probability flow determined by the
classical Hamilton-Jacobi equation (\ref{classical HJ}). Therefore for
$\alpha^{\prime}\rightarrow\infty$ the particles follow classical
trajectories. We conclude that the $\alpha^{\prime}\rightarrow\infty$ limit of
the hybrid theory is classical mechanics.

Here too a word of caution is needed. We say the emergent entropic dynamics in
the $\xi=0$ case is \textquotedblleft essentially\textquotedblright\ classical
mechanics. The point is that here too, entropic dynamics remains
\textquotedblleft entropic\textquotedblright. Even for very large
$\alpha^{\prime}$, at sufficiently microscopic scales the expected motion
remains Brownian.

To summarize: Within the universality class of quantum dynamics, suppressing
microscopic fluctuations yields Bohmian mechanics. Similarly, within the
universality class of hybrid dynamics, suppressing microscopic fluctuations
yields classical mechanics.

\section{Momentum and its uncertainty relations}

In the ED of particles there is one set of beables --- their positions. This
immediately raises questions about the nature of other observables: are they
beables or are they created in the act of measurement? The case of momentum
and the Heisenberg uncertainty relation is discussed in \cite{Nawaz Caticha
2011}; other related matters in \cite{Caticha 2012} and \cite{Johnson Caticha
2011}.

In this paper we have seen that ED allows the construction of a variety of
models that can differ both at the microscopic and at the macroscopic level
--- we can enhance or suppress microscopic fluctuations by tuning
$\alpha^{\prime}$ and we can turn the quantum potential on and off by setting
$\xi>0$ or $\xi=0$. Our goal here is to revisit the issue of momentum and its
uncertainty relations to find out how they are affected by the choices of
$\alpha^{\prime}$ and $\xi$.

First, let us recall the notion of momentum. For simplicity we consider a
single particle. Since it follows a non-differentiable trajectory it is clear
that the classical momentum $md\vec{x}/dt$ tangent to the trajectory cannot be
defined. One can however introduce two notions of momentum. One is the usual
quantum operator,
\begin{equation}
\hat{p}^{a}=-i\hbar\delta^{ab}\partial_{b}~, \label{q_m}%
\end{equation}
which acts on the space of wave functions $\Psi$ and generates infinitesimal
translations. The other momentum is a local quantity associated to the current
velocity, $v^{A}=m^{AB}\partial_{B}\Phi$ given in eq.(\ref{phase}), which
leads to
\begin{equation}
p_{A}(x)=m_{AB}v^{B}(x)=\partial_{A}\Phi(x)~.
\end{equation}
For a single particle this \textquotedblleft local\textquotedblright\ or
\textquotedblleft current\textquotedblright\ momentum is
\begin{equation}
p^{a}(x)=mv^{a}(x)=\delta^{ab}\partial_{b}\Phi(x)~. \label{cur_m}%
\end{equation}
As we see from (\ref{phase}), the local momentum can be decomposed into drift
and osmotic components, $p^{a}=p_{d}^{a}+p_{o}^{a}$ where
\begin{equation}
p_{d}^{a}=mb^{a}=\delta^{ab}\tilde{\eta}\partial_{b}\phi~, \label{drift_m}%
\end{equation}%
\begin{equation}
p_{o}^{a}=mu^{a}=-\delta^{ab}\frac{\tilde{\eta}}{\alpha^{\prime}}\partial
_{b}\log\rho^{1/2}\,. \label{oss_m}%
\end{equation}

Two points deserve to be emphasized. First, the two notions of momentum,
(\ref{q_m}) and (\ref{cur_m}), are not unrelated: both generate displacements.
As shown in \cite{Caticha 2015} the infinitesimal displacement of a functional
$f[\rho,\Phi]$ is given by its Poisson bracket with the \textquotedblleft
ensemble\textquotedblright\ momentum
\begin{equation}
\tilde{P}^{a}[\rho,\Phi]=\int d^{3}x\,\rho\delta^{ab}\partial_{b}\Phi=\langle
p^{a}\rangle~.
\end{equation}
The second point is that the local momentum $p^{a}(x)$ is\ expressed in terms
of the probability $\rho$ and the drift potential $\phi$. It is not a beable;
it is an epistemic concept. \emph{It is not an attribute of the particle but
of the wave function}.

Next we revisit the uncertainty relation. We recall that the relation between
expected values follows from
\begin{equation}
\langle\partial_{a}\log\rho\rangle=0\Rightarrow\langle p_{o}^{a}\rangle
=0\quad\text{and}\quad\langle p^{a}\rangle=\langle p_{d}^{a}\rangle~.
\end{equation}
Also, using $\Psi=\rho^{1/2}\exp(i\Phi/\hbar)$, we have
\begin{equation}
\langle\hat{p}^{a}\rangle=\int dx\,\Psi^{\ast}\frac{\hbar}{i}\delta
^{ab}\partial_{b}\Psi=\langle p^{a}\rangle~.\label{q_ave}%
\end{equation}
The corresponding covariances are related by
\begin{equation}
\text{Cov}(\hat{p}^{a},\hat{p}^{b})=\langle\hat{p}^{a}\hat{p}^{b}%
\rangle-\langle\hat{p}^{a}\rangle\langle\hat{p}^{b}\rangle=\text{Cov}%
(p^{a},p^{b})+\frac{\hbar^{2}}{4}I^{ab}~,\label{cov_pq}%
\end{equation}
where $I^{ab}=\delta^{ac}\delta^{bd}I_{cd}\ $and
\begin{equation}
I_{cd}=\int dx\,\rho\partial_{c}\log\rho\partial_{d}\log\rho=\text{Cov}%
(\partial_{c}\log\rho,\partial_{d}\log\rho)~,\label{Fisher info}%
\end{equation}
is the Fisher information matrix. The variances of $\hat{p}^{a}$ and $p^{a}$
are related by
\begin{equation}
\text{Cov}(\hat{p}^{a},\hat{p}^{a})=\text{Var}(\hat{p}^{a})=\text{Var}%
(p^{a})+\frac{\hbar^{2}}{4}\text{Var}(\partial^{a}\log\rho)~.\label{v_q}%
\end{equation}
(No sum on repeated $a$'s.) The uncertainty relation is obtained by
multiplying (\ref{v_q}) by Var$(x^{b})$,
\begin{equation}
\text{Var}(\hat{p}_{q}^{a})\text{Var}(x^{b})=\text{Var}(p^{a})\text{Var}%
(x^{b})+\frac{\hbar^{2}}{4}\text{Var}(\partial^{a}\log\rho)\,\text{Var}%
(x^{b})~.
\end{equation}
and invoking the Cauchy-Schwartz inequality,
\begin{equation}
\text{Var}(A)\text{Var}(B)\geq\text{Cov}^{2}(A,B)~,\label{cs}%
\end{equation}
to get
\begin{equation}
\text{Var}(\hat{p}^{a})\text{Var}(x^{b})\geq\text{Cov}^{2}(p^{a},x^{b}%
)+\frac{\hbar^{2}}{4}\text{Cov}^{2}(\partial^{a}\log\rho,x^{b})~.\label{ur a}%
\end{equation}
Finally, an explicit calculation of the covariances on the right gives
\begin{equation}
\text{Cov}(\hat{p}^{a},x^{b})=\frac{1}{2}\langle\hat{p}^{a}\hat{x}^{b}+\hat
{x}^{b}\hat{p}^{a}\rangle-\langle\hat{p}^{a}\rangle\langle\hat{x}^{b}%
\rangle=\text{Cov}(p^{a},x^{b})~,\label{cov_a}%
\end{equation}
and
\begin{equation}
\text{Cov}(\partial^{a}\log\rho,x^{b})=-\delta^{ab}.\label{cov_b}%
\end{equation}
Substituting into (\ref{ur a}) leads to
\begin{equation}
\text{Var}(\hat{p}^{a})\text{Var}(x^{b})\geq\text{Cov}^{2}(\hat{p}_{q}%
^{a},x^{b})+\left(  \frac{\hbar}{2}\right)  ^{2},\label{var_q1}%
\end{equation}
which is the version of the uncertainty relation proposed originally by
Schr\"{o}dinger. The weaker version due to Heisenberg follows immediately,
\begin{equation}
\text{Var}(\hat{p}^{a})\text{Var}(x^{b})\geq\left(  \frac{\hbar}{2}\right)
^{2}\quad\text{or}\quad\Delta\hat{p}^{a}\Delta x^{b}\geq\frac{\hbar}%
{2}~.\label{hup}%
\end{equation}

The derivation above shows that we can switch the quantum potential off by
setting $\xi=0$ without affecting either the momentum or the uncertainty
relation. It also shows that the value of $\alpha^{\prime}$ never entered the
argument. We can enhance or suppress microscopic fluctuations without
affecting the momentum or the uncertainty relation. $\alpha^{\prime}$ can
affect the relative drift and osmotic components of the local momentum, but
not the local momentum itself.

\paragraph{Acknowledgments}

We would like to thank M. Abedi, C. Cafaro, N. Caticha, S. DiFranzo, A.
Giffin, S. Ipek, D.T. Johnson, K. Knuth, S. Nawaz, C. Rodr\'{\i}guez, K.
Vanslette, and very specially M. Reginatto for many discussions on entropy,
inference and quantum mechanics.


\begin{thebibliography}{99}                                                                                               %


\bibitem {Caticha 2012}For a pedagogical review see A. Caticha, \emph{Entropic
Inference and the Foundations of Physics} (monograph commissioned by the 11th
Brazilian Meeting on Bayesian Statistics -- EBEB-2012); http://www.albany.edu/physics/ACaticha-EIFP-book.pdf.

\bibitem {Caticha 2010a}A. Caticha, J. Phys. A: Math. Theor. \textbf{2011,
}44, 225303; arXiv.org:1005.2357.

\bibitem {Caticha 2014a}A. Caticha, J. Phys.: Conf. Ser. \textbf{2014}, 504,
012009; arXiv.org:1403.3822.

\bibitem {Caticha 2014b}A. Caticha, D. Bartolomeo, M. Reginatto, in
\emph{Bayesian Inference and Maximum Entropy Methods in Science and
Engineering}, ed. by A. Mohammad-Djafari, and F. Barbaresco, AIP Conf. Proc.
\textbf{1641}, 155 (2015); arXiv.org:1412.5629.

\bibitem {Caticha 2015}A. Caticha, Entropy \textbf{17}, 6110 (2015); arXiv.org:1509.03222.

\bibitem {Leifer 2014}M. S. Leifer \textquotedblleft Is the Quantum State
Real? An Extended Review of $\psi$-ontology Theorems\textquotedblright, Quanta
\textbf{3}, 67--155 (2014); arXiv:1409.1570.

\bibitem {Wootters 1981}W. K. Wootters, Phys. Rev. D \textbf{23}, 357 (1981).

\bibitem {Caticha 1998}A. Caticha, Phys. Lett. A \textbf{244}, 13 (1998);
Phys. Rev. A \textbf{57}, 1572 (1998); Found. Phys. \textbf{30}, 227 (2000).

\bibitem {Bruckner Zeilinger 2002}C. Brukner, A. Zeilinger, \textquotedblleft
Information and Fundamental Elements of the Structure of Quantum
Theory\textquotedblright\ in\ \emph{Time, Quantum, Information}, ed. by L.
Castell and O. Ischebeck (Springer 2003); arXiv:quant-ph/0212084.

\bibitem {Spekkens 2007}R. Spekkens, Phys. Rev. A \textbf{75}, 032110 (2007).

\bibitem {Goyal Knuth Skilling 2010}P. Goyal, K. Knuth, and J. Skilling, Phys.
Rev. A \textbf{81}, 022109 (2010).

\bibitem {Hardy 2011}L. Hardy, \textquotedblleft Reformulating and
Reconstructing Quantum Theory\textquotedblright, arXiv.org:1104.2066.

\bibitem {Hall Reginatto 2002}M.J.W. Hall and M. Reginatto, J. Phys. A
\textbf{35}, 3289 (2002); Fortschr. Phys. \textbf{50}, 646 (2002); arXiv:quant-ph/0201084.

\bibitem {Reginatto 2013}M. Reginatto, \textquotedblleft From information to
quanta: a derivation of the geometric formulation of quantum theory from
information geometry\textquotedblright, arxiv:1312.0429.

\bibitem {Nelson 1985}Nelson, E. \emph{Quantum Fluctuations} (Princeton UP,
Princeton 1985).

\bibitem {de la Pena et al 2015}L. de la Pe\~{n}a, A. M. Cetto, A.
Valdez-Hern\'{a}ndez, \textquotedblleft\emph{The Emerging Quantum: the Physics
Behind Quantum Mechanics}\textquotedblright\ (Springer, 2015).

\bibitem {Smolin 2006}L. Smolin, \textquotedblleft Could quantum mechanics be
an approximation to another theory?\textquotedblright\ arXiv.org/abs/quant-ph/0609109.

\bibitem {tHooft 2002}G. 't Hooft, \textquotedblleft Determinism beneath
Quantum Mechanics\textquotedblright, arxiv:quant-ph/0212095; \textquotedblleft
Emergent quantum mechanics and emergent symmetries\textquotedblright, arxiv:hep-th/0707.4568.

\bibitem {Elze 2002}H.T. Elze and O. Schipper, Phys. Rev. D \textbf{66},
044020 (2002); H.T. Elze, Phys. Lett. A \textbf{310}, 110 (2003).

\bibitem {Adler 2004}S. Adler, \emph{Quantum Theory as an Emergent Phenomenon}
(Cambridge UP, Cambridge 2004).

\bibitem {Grossing 2008}G. Gr\"{o}ssing, Phys. Lett. A \textbf{372}, 4556
(2008), arxiv:0711.4954; G. Gr\"{o}ssing \emph{et al}, J. Phys.: Conf. Ser.
\textbf{361}, 012008 (2012).

\bibitem {Bartolomeo Caticha 2015}Some of these ideas have been briefly
reported in D. Bartolomeo and A. Caticha, \textquotedblleft Entropic Dynamics:
The Schr\"{o}dinger equation and its Bohmian limit\textquotedblright, arXiv.org:1512.09084.

\bibitem {Bohm 1952}D. Bohm, Phys. Rev. \textbf{85}, 166 and 180 (1952).

\bibitem {Bohm Hiley 1993}D. Bohm and B. J. Hiley, \emph{The undivided
universe -- An ontological interpretation of quantum theory} (Routlege, New
York, 1993).

\bibitem {Holland 1993}P. R. Holland, The Quantum Theory of Motion (Cambridge
UP, 1993).

\bibitem {Nawaz Caticha 2011}S. Nawaz and A. Caticha, in\ \emph{Bayesian
Inference and Maximum Entropy Methods in Science and Engineering}, ed. by K.
Knuth \emph{et al.}, AIP Conf. Proc. \textbf{1443}, 112 (2012); arXiv:1108.2629.

\bibitem {de Falco et al 1982}D. de Falco, S. D. Martino and S. De Siena,
Phys. Rev. Lett. \textbf{49}, 181 (1982).

\bibitem {De Martino et al 1984}S. De Martino and S. De Siena, Nuovo Cimento
\textbf{79B}, 175 (1984).

\bibitem {Golin 1985}S. Golin, J. Math. Phys. \textbf{26}, 2781 (1985).

\bibitem {Golin 1986}S. Golin, J. Math. Phys. \textbf{27}, 1549 (1986).

\bibitem {Johnson Caticha 2011}D. T. Johnson and A. Caticha,
in\ \emph{Bayesian Inference and Maximum Entropy Methods in Science and
Engineering}, ed. by K. Knuth \emph{et al.}, AIP Conf. Proc. \textbf{1443},
104 (2012); arXiv:1108.2550.
\end{thebibliography}
\end{document}